\documentclass[12pt,preprint]{aastex}
\usepackage{color,verbatim} 

\usepackage{epsfig}
\usepackage{subfigure}
\usepackage{graphicx}
\usepackage{amsmath}
\usepackage{natbib}

\newcommand{\mb}{\boldsymbol}

\shorttitle{Planetesimal Formation} \shortauthors{Bai \&
Stone}

\begin{document}


\title{The Effect of the Radial Pressure Gradient in
Protoplanetary Disks on Planetesimal Formation}


\author{Xue-Ning Bai \& James M. Stone}
\affil{Department of Astrophysical Sciences, Princeton University,
Princeton, NJ, 08544} \email{xbai@astro.princeton.edu,
jstone@astro.princeton.edu}




\begin{abstract}
The streaming instability (SI) provides a promising mechanism for
planetesimal formation because of its ability to concentrate solids into
dense clumps. The degree of clumping strongly depends on the
height-integrated solid to gas mass ratio $Z$ in protoplanetary disks (PPDs).
In this letter, we show that the magnitude of the radial pressure gradient
(RPG) which drives the SI (characterized by $q\equiv\eta v_K/c_s$, where
$\eta v_K$ is the reduction of Keplerian velocity due to the RPG and $c_s$
is the sound speed) also
strongly affects clumping. We
present local two-dimensional hybrid numerical simulations of
aerodynamically coupled particles and gas in the midplane of PPDs.
Magnetic fields and particle self-gravity are ignored.  We explore three
different RPG values appropriate for typical PPDs: $q=0.025, 0.05$ and $0.1$.
For each $q$ value, we consider four different particle size
distributions ranging from sub-millimeter to meter sizes and run
simulations with solid abundance from $Z=0.01$ up to $Z=0.07$. We
find that a small RPG strongly promotes particle clumping in that:
1) At fixed particle
size distribution, the critical solid abundance $Z_{\rm crit}$ above which
particle clumping occurs monotonically increases with $q$; 2) At fixed $Z$,
strong clumping can occur for smaller particles when $q$ is smaller.
Therefore, we expect planetesimals to form preferentially in regions
of PPDs with a small RPG.
\end{abstract}


\keywords{diffusion --- hydrodynamics --- instabilities --- planetary systems:
protoplanetary disks --- planets and satellites: formation --- turbulence}

\section{Introduction}\label{sec:intro}

Planetesimals are super-kilometer sized bodies that are the building
blocks of planets \citep{Safranov69,ChiangYoudin10}, yet their
formation has long been a mystery. Millimeter and centimeter sized
particles are routinely observed in protoplanetary disks (PPDs)
\citep{Wilner_etal05,Rodmann_etal06,Natta_etal07,Lommen_etal10},
but particles with larger size seem difficult to form by coagulation
\citep{BlumWurm08,Guttler_etal10}. Solids close to meter
size further suffer from rapid radial drift due to the negative radial pressure
gradient (RPG) in PPDs \citep{Weidenschilling77}. The gravitational
instability (GI) scenario of planetesimal formation \citep{GW73} also has
difficulties, because even without an external source of turbulence, the
Kelvin-Helmholtz instability (KHI) generated from the dusty midplane layer
prevents the onset of GI unless the local height-integrated solid to gas
mass ratio ($Z$, hereafter referred to as solid abundance) is about an
order of magnitude above solar metallicity 
\citep{Weidenschilling80,Sekiya98,YoudinShu02}.

It was found recently that the drag interaction between solids and gas
leads to a powerful instability \citep{GoodmanPindor00}. This  ``streaming
instability" (SI, \citealp{YoudinGoodman05}) results in spontaneous particle
clumping from the equilibrium state between solids and gas \citep{NSH86}.
Numerical simulations demonstrate that the non-linear saturation of the SI
concentrates particles into dense clumps \citep{JohansenYoudin07},
promoting planetesimal formation by collective particle self-gravity,
bypassing the meter-size barrier. Indeed,
\cite{Johansen_etal07,Johansen_etal09} showed that planetesimals 
with sizes of a few hundred kilometers
form rapidly by the SI from centi-meter to decimeter sized pebbles and rocks,
consistent
with constraints from the asteroid belt \citep{Morbidelli_etal09}.

This letter complements our previous work on the dynamics of particle and
gas in the midplane of PPDs (\citealp{BaiStone10b}, hereafter BS10).
We consider a wide size distribution of particles ranging from
sub-millimeter up to meter sizes as an approximation for the outcome of
dust coagulation in PPDs \citep{Birnstiel_etal10,Zsom_etal10}. External
sources of turbulence such as magnetorotational instability are ignored, as
appropriate for the dead zones in PPDs
\citep{Gammie96,Stone_etal00,BaiGoodman09},
while SI is the main source of turbulence due to particle settling.
Particle self-gravity is ignored in our simulations, as we focus on the
precursor of planetesimal formation: particle clumping.


The SI is powered by the RPG in the gaseous disk. The RPG reduces the
gas orbital velocity (in the absence of solids) by a fraction $\eta$ of the
Keplerian velocity $v_K=\Omega r$, and is characterized by
\begin{equation}
q\equiv\eta v_K/c_s=\eta r/H_g\ ,\label{eq:qdef}
\end{equation}
where $c_s$ is the isothermal sound speed, and $H_g=c_s/\Omega$ is the
scale height of the gaseous disk. In BS10, we assumed $q=0.05$ throughout,
while in reality $q$ depends on the parameters of, and location in, the
disk.  The strength of the SI turbulence scales with
the RPG, which in turn affects the particle-gas dynamics in the disk midplane.
In particular, we show in this letter that particle clumping strongly depends
on the RPG, which has important implications for planetesimal formation.

\section[]{Simulations}\label{sec:setup}

\begin{table}
\caption{Run parameters.}\label{tab:simulation}
\begin{center}
\begin{tabular}{cccccccccccc}\hline\hline
 Run & $q$ & $100Z$ & $\tau_{\rm min}$ & $\tau_{\rm max}$ & $L_x\times L_z\ ^1$
 & $N_x\times N_z\ ^2$ \\\hline
        & 0.025 & 1..3 &  &  & $0.05\times0.15$ & $256\times768$ \\
 R41 & 0.05   & 1..5 & $10^{-4}$ & $10^{-1}$ & $0.1\times0.3$ & $256\times768$ \\ 
         & 0.1    & 1..7 &  &  & $0.2\times0.4$ & $256\times512$ \\\hline
         
        & 0.025 & 1..3 &  &  & $0.05\times0.15$ & $256\times768$ \\
 R21 & 0.05   & 1..4 & $10^{-2}$ & $10^{-1}$ & $0.1\times0.3$ & $256\times768$ \\
         & 0.1    & 1..7 &  &  & $0.2\times0.4$ & $256\times512$ \\\hline
         
         & 0.025 & 1..3 &  &  & $0.2\times0.3$ & $256\times512$ \\
 R30 & 0.05   & 1..3 & $10^{-3}$ & $1$      & $0.2\times0.3$ & $256\times384$ \\
         & 0.1     & 1..7 &  &  & $0.2\times0.3$ & $256\times384$ \\\hline
         
         & 0.025 & 1..3 &   &  & $0.2\times0.3$ & $256\times512$ \\
 R10 & 0.05   & 1..3 & $10^{-1}$ & $1$       & $0.2\times0.3$ & $256\times384$ \\
         & 0.1     & 1..6 &  &  & $0.2\times0.3$ & $256\times384$ \\
\hline\hline
\end{tabular}
\end{center}

$^1$ Domain size, in unit of gas scale height $H_g=c_s/\Omega$.

$^2$ Grid resolution.
\end{table}

We perform two-dimensional (2D) hybrid simulations of gas and solids using
the Athena code \citep{AthenaTech,BaiStone10a}, where gas is treated as
a hydrodynamical fluid (without magnetic field) on an Eulerian grid,
and the solids
are treated as superparticles, each representing a swarm of real particles.
We model a local patch of the PPDs using the shearing sheet approximation.
The dynamical equations are written in a reference frame corotating at the
Keplerian frequency $\Omega$ at fiducial radius $r$. We assume
axisymmetry, and the simulations are performed in the radial-vertical
($x$-$z$) plane, where ${\mb\Omega}$ is along the $z$ direction and $x$
points radially outward. The particles are coupled to the gas via
aerodynamic drag, characterized by the stopping time $t_{\rm stop}$, with
momentum feedback included. The gas is assumed to be isothermal, with
midplane density $\rho_g$. Vertical gravity $g_z=-\Omega^2z$ is included for
both particles and the gas.


The dynamical equations and simulation setup are identical to those in
BS10. Specifically, we consider a particle size distribution that
is discretized into a number of particle size bins, with each bin covering half a
dex in $\tau_s\equiv\Omega t_{\rm stop}$, bounded by the minimum and
maximum stopping time $\tau_{\rm min}$ and $\tau_{\rm max}$.  In what follows,
we label our
simulations with names of the form R$mn$, where $m,n$ are integers
obtained by $\tau_{\rm min}=10^{-m}$ and $\tau_{\rm max}=10^{-n}$, thus
run R$mn$ uses $2(m-n)+1$ particle size bins (or particle species). We use
a variety of grid resolution and box sizes to capture the fastest growing
modes of the SI, and we use $10^5$
particles per species in all our simulations. We assume uniform mass
distribution across all particle size bins, with total solid abundance $Z$.
As in BS10, we consider four groups of runs, R41, R21, R30 and R10. 

For each particle size distribution, we perform simulations with three different
values of the RPG parameter: $q=0.025, 0.05$ and $0.10$. According to
equations (4) and (7) in BS10, the dependence of $q$ on disk parameters such
as temperature and mass
is relatively weak, thus our range of $q$ covers a large parameter
space of disk models. For each set of runs, we perform a series of simulations
with different $Z$ values, starting at $Z=0.01$ and increasing the value
by $0.01$ for each new run, until strong particle clumping occurs or $Z=0.07$.  Our
simulation run parameters are summarized in Table \ref{tab:simulation}. They
are identical to the 2D run parameters in BS10, but use different
values for $q$ and a larger range in $Z$. Since the natural length scale
of the SI is $\eta r$, we use smaller (bigger) simulation box sizes
for smaller (larger)
$q$ values so that $\eta r$ is resolved by an equal number of grid cells in each
series of runs. All simulations are run for at least $900\Omega^{-1}$.

Three-dimensional (3D) simulation with the inclusion of the azimuthal
dimension is necessary to capture the KHI \citep{Chiang08,Barranco09},
which is mainly caused by the vertical shear in the gas azimuthal velocity.
Our simulations are 2D rather than 3D for two reasons. First, for our adopted
particle size distribution, the turbulence generated from the SI stops particles
from settling before the onset KHI (BS10), at least for the
$q=0.05$ case. Second, in order to properly resolve the SI, relatively high
resolution is required \citep{BaiStone10a}, and particle clumping
does depend on resolution\footnote{For example, Run R30Z3-3D in BS10
has no particle clumping in our standard resolution, but shows clumping
at lower resolution.}. Simulations in 3D with such high resolution are
too costly due to the small box size (thus time step),
especially for $q=0.025$.

\section[]{Particle Clumping}\label{sec:clumping}

\begin{figure*}
    \centering
    \includegraphics[width=180mm,height=60mm]{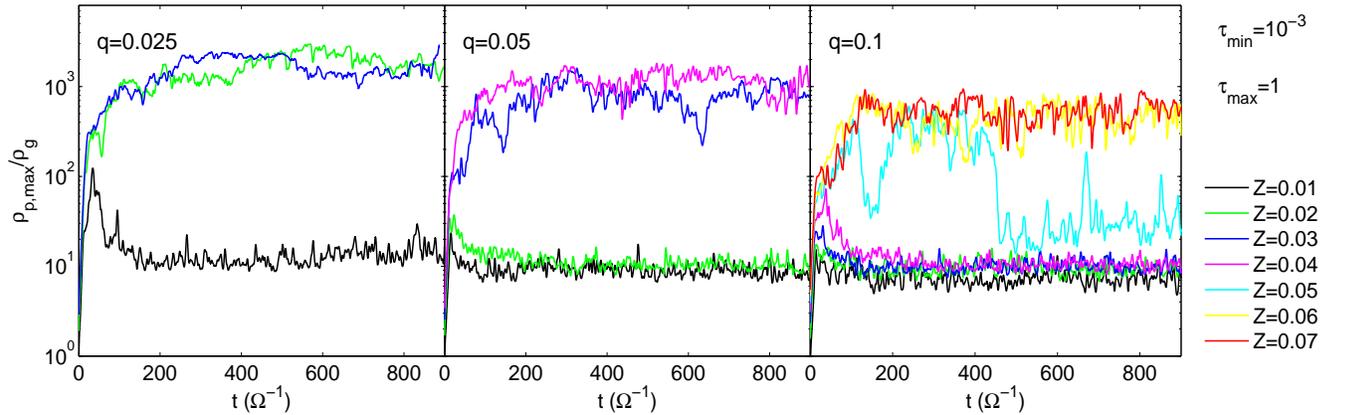}
  \caption{Time history of the maximum particle density in simulations
with a range of particle sizes from $\tau_s = 10^{-3}$ to one (runs
R30). Left, middle and right panels show the results from $q=0.025, 0.05$
  and $0.1$ respectively. Runs with different solid abundance $Z$ are
  labeled with different colors.}\label{fig:history}
\end{figure*}

The simulations saturate in about 50-100 orbits. Particle settling triggers
the SI, and particles with different stopping times are maintained at different
heights determined by the balance between settling and turbulent diffusion.
Most interestingly, SI efficiently concentrate particles into dense clumps when
the solid abundance is sufficiently high (\citealp{Johansen_etal09}, BS10).
In Figure \ref{fig:history} we plot the evolution of maximum particle density
$\rho_{p,{\rm max}}$ for the R30 runs with different values of $q$ and $Z$.
There is a clear dichotomy in the evolution: either the maximum particle
density stays at a
relatively small value ($\rho_{\rm p,max}\lesssim50\rho_g$), or it reaches as
high as $10^3\rho_g$, indicative of strong clumping. A particle clump
becomes gravitationally bound when its density exceeds the Roche density
(see equation (18) of BS10), which also is of the order $10^3\rho_g$ for
typical PPDs. Therefore, particle clumping is a prelude to planetesimal formation
via gravitational collapse. For each value of $q$, the transition from
non-clumping to clumping is sharp as $Z$ gradually increases, and
we can define a critical solid abunance $Z_{\rm crit}$, where strong particle
clumping occurs for $Z>Z_{\rm crit}$. Comparing different panels indicates
that $Z_{\rm crit}$ monotonically increases with $q$.

\begin{figure*}
    \centering
    \includegraphics[width=180mm,height=60mm]{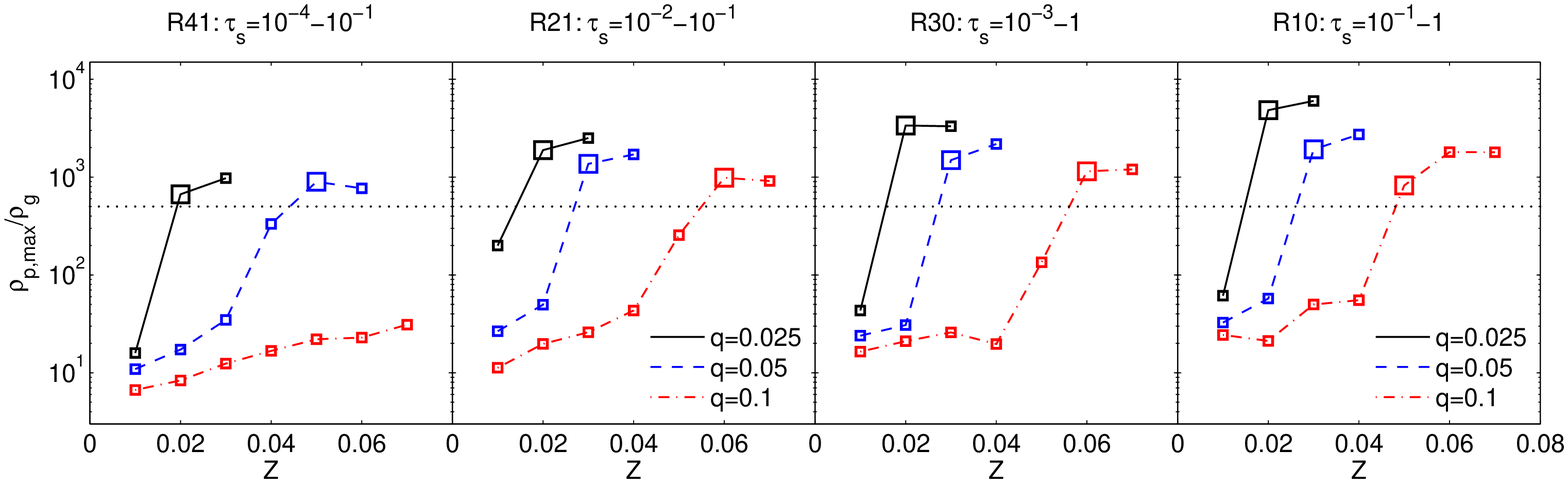}
  \caption{The maximum particle density that can be achieved by SI from
  all our simulation runs (for R41, R21, R30, R10 from left to right panels).
  In each panel, we plot the maximum particle density as a function of solid
  abundance $Z$ for each value of $q$ (black solid, blue dashed and red
  dash-dotted for $q=0.025, 0.5$ and $0.1$ respectively). The dotted line
  at $\rho_{p,{\rm max}}=500\rho_g$ indicates the adopted threshold for
  strong particle clumping, and we use larger symbols to indicate the
  {\it first} runs that show strong clumping as $Z$ increases.}\label{fig:RPG}
\end{figure*}

To better demonstrate the effect of the disk RPG on planetesimal formation, we
plot the maximum particle density as a function of solid abundance $Z$ for
each group of simulations with all values of $q$ in Figure \ref{fig:RPG}. The
maximum density is taken from the largest value of $\rho_{p,{\rm max}}$
over the last $20$ orbits of the simulations, when all the runs are fully
saturated. We take $\rho_p=500\rho_g$ (which is the same order as the Roche
density) as a rough indicator of planetesimal formation, as shown by the
dotted lines. For each value of $q$, one can determine $Z_{\rm crit}$ at the
intersection of the dashed and corresponding solid lines. We see that
for all four groups of runs, $Z_{\rm crit}$ monotonically increases with $q$.
Moreover, $Z_{\rm crit}$ depends on $q$ more sensitively when the particle
size is on average smaller. When $q=0.025$, $Z_{\rm crit}$ is about $0.015$
for all our four groups of runs. For $q=0.1$, run R41 does not show particle
clumping below $Z=0.07$, for runs R21 and R30 $Z_{\rm crit}$ is about
$0.06$, while for runs R10 $Z_{\rm crit}$ drops to below $0.05$.

Particle clumping is a highly non-linear effect due to the SI, and is more likely
to develop when the average disk midplane solid to gas mass ratio $\epsilon$
is large. If one assumes $D$ to be the midplane vertical diffusion coefficient
of the SI turbulence, particles with stopping time $\tau_s$ would settle
to a layer with thickness of the order $H_p\approx\sqrt{D/\Omega\tau_s}$. The
diffusion coefficient $D$ depends on the particle size distribution,
$\epsilon\approx ZH_g/H_p$ and $q$. Without vertical gravity, the only length
scale in the problem is $\eta r=qH_g$, thus $D\propto q^2$. If one assumes
$D\propto\epsilon^\alpha$, we obtain by generalizing the toy model
in BS10 (see their equation (16) and (17))
\begin{equation}
H_p\propto Z^{\alpha/(\alpha+2)}q^{2/(\alpha+2)}\ .
\end{equation}
Since one expects $\alpha<0$ for relatively large $\epsilon\gtrsim1$, we see that
$H_p$ not only depends sensitively on $Z$, as shown in BS10, it depends even
more sensitively on $q$. This critical dependence makes the average particle
to gas mass ratio $\epsilon$ at midplane quickly increases as $q$ decreases,
promoting strong particle clumping, which explains the trend in Figure
\ref{fig:RPG}. The more sensitive dependence of $Z_{\rm crit}$ on $q$ for
smaller particles can be interpreted as a result of the power-law index $\alpha$
tending to be more negative for smaller $\tau_s$.

\begin{figure*}
    \centering
    \includegraphics[width=180mm,height=60mm]{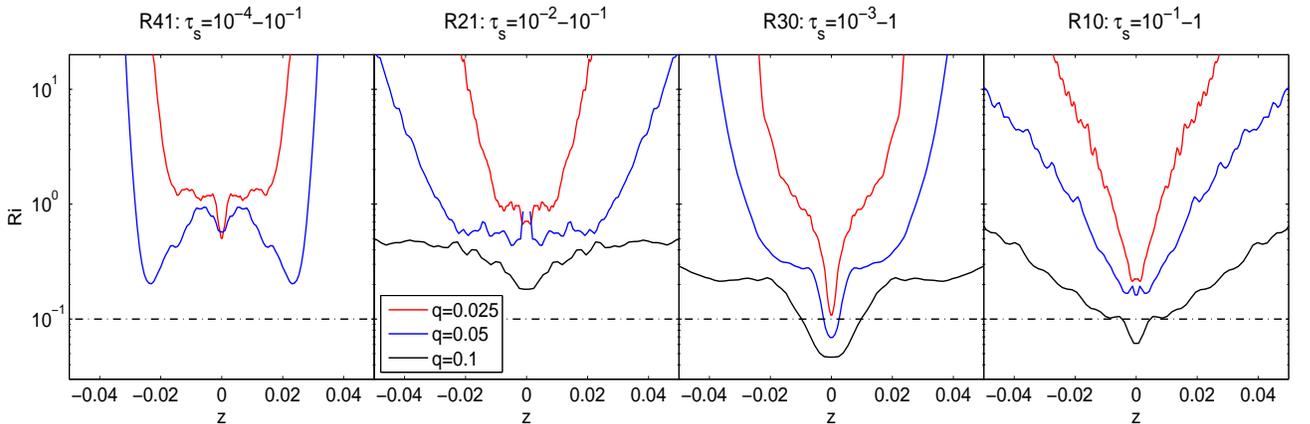}
  \caption{The Richardson number profile for all simulation runs (for R41, R21, R30,
  R10 from left to right panels) with solid abundance just above $Z_{\rm crit}$ (the
  runs with enlarged symbols in Figure \ref{fig:RPG}). Results from $q=0.025, 0.05$
  and $0.1$ are shown in black solid, blue dashed and red dash-dotted curves
  respectively. The dotted line indicated an approximate estimate of the critical
  Richardson number $Ri_{\rm crit}=0.1$.}\label{fig:richardson}
\end{figure*}

Our 2D simulations do not contain the azimuthal dimension, which
suppresses KHI.  Nevertheless, we can check whether KHI would occur
before particles settle into a sufficiently thin layer to trigger
strong SI by plotting the vertical profile of the Richardson number for
simulations at critical solid abundance for all $q$ values, see Figure
\ref{fig:richardson}.  The Richardson number is calculated by the method
described in \S3.1 of BS10. Although the Richardson number alone does not
determine the stability when Coriolis force \citep{GomezOstriker05}, and
radial shear \citep{Barranco09} are present, one may still take $Ri_{\rm
crit}=0.1$ as an approximation for the critical value for instability
\citep{Chiang08,Lee_etal10}.  We see in Figure \ref{fig:richardson} that
for most of these runs, the Richardson number across the disk is above
the critical value, meaning that the system is KH stable at $Z_{\rm
crit}$. Moreover, as $q$ increases, the Richardson number at $Z_{\rm
crit}$ decreases. In particular, for R30 and R10 runs with $q=0.1$, the
Richardson number at the disk midplane drops below $0.1$, which may be
subject to KHI. Therefore, the condition for strong clumping for these
runs may be more stringent than shown in Figure \ref{fig:RPG}.

\section[]{Discussions and Conclusions}

In this letter, we have demonstrated that a small RPG strongly favors particle
clumping and planetesimal formation via three effects. First, for a fixed
particle size distribution, the critical solid abundance $Z_{\rm crit}$ above
which strong clumping occurs is reduced. Second, at fixed solid abundance,
strong clumping occurs for smaller particles. Third, KHI is less likely to be
triggered (and therefore will not suppress particle clumping)
at $Z=Z_{\rm crit}$. Moreover, a
smaller RPG also implies smaller radial drift and collision velocities,
promoting grain growth, strengthening our second point above. A small RPG
also favors the GI scenario of planetesimal formation (relevant when all
particles are strongly coupled to the gas) because there is less free energy
available in the vertical shear \footnote{We expect the threshold abundance
for GI to operate is larger than $Z_{\rm crit}$.}. 

One caveat from our 2D simulations is that the condition for particle clumping
is slightly more stringent in 3D than in 2D (BS10). Therefore, the values of
$Z_{\rm crit}$ obtained in this letter may be considered as a lower bound.
Nevertheless, the conclusions in this letter are robust. In fact, Johansen et al.
(2007, see their supplemental material) also found that planetesimal formation
is eased with smaller pressure gradient using 3D simulations.

Combined with the results in BS10, we have shown that strong particle
clumping is favored when there is: 1) a small RPG; 2) large solid abundance;
3) large solid size; and 4) a dead zone. These results indicate that planetesimals
preferentially form in specific locations in PPDs. The global structure and
evolution of PPDs is of crucial importance: it determines the RPG
profile, the radial transport of particles which leads to enhancement of
solid abundance in the inner disk \citep{YoudinShu02}, and grain
coagulation that determines the particle size distribution. Recent models on
the structure and evolution of PPDs (e.g., \citealp{JinSui10,Zhu_etal10b})
generally show much more complicated  structures than the simple MMSN
or $\alpha$-disk models due to non-steady state accretion as well as the
presence of dead zones. In addition, RPG may reach zero in local pressure
bumps at the snow line \citep{KretkeLin07} or the inner edge of the dead
zone \citep{Dzyurkevich_etal10} which can also be the preferred site for
planetesimal formation.  These results all-together provide useful constraints
on the initial conditions for the formation of planet embryos and planets
\citep{KokuboIda00,KokuboIda02}.

\acknowledgments

We are grateful to D.N.C. Lin for helpful discussions. This work is supported
by NSF grant AST-0908269. XNB acknowledges support from NASA Earth
and Space Science Fellowship.


\label{lastpage}
\end{document}